# High temperature, gate-free quantum anomalous Hall effect with an active capping layer

Hee Taek Yi[1,2], Deepti Jain[1], Xiong Yao[2]†, and Seongshik Oh[1,2]*

[1] Department of Physics and Astronomy, Rutgers, The State University of New Jersey, Piscataway, NJ 08854, USA.
[2] Center for Quantum Materials Synthesis, Piscataway, NJ 08854, USA.
†Present address, Ningbo Institute of Materials Technology and Engineering, Chinese Academy of Sciences, Ningbo 315201, China.
*Corresponding author. Email: ohsean@physics.rutgers.edu

**Quantum anomalous Hall effect (QAHE) was discovered a decade ago, but is still not utilized beyond a handful of research groups, due to numerous limitations such as extremely low temperature, electric field-effect gating requirement, small sample sizes and environmental aging effect. Here, we present a robust platform that provides effective solutions to these problems. Specifically, on this platform, we observe QAH signatures at record high temperatures, with the Hall conductance of 1.00 $e^2/h$ at 2.0 K, 0.98 $e^2/h$ at 4.2 K, and 0.92 $e^2/h$ at 10 K, on centimeter-scale substrates, without electric-field-effect gating. The key ingredient is an active $CrO_x$ capping layer, which substantially boosts the ferromagnetism while suppressing environmental degradation. With this development, QAHE will now be accessible to much broader applications than before.**

Although the possibility of QAHE, which exhibits quantized transverse (Hall) conductance without an external magnetic field,[1, 2] was first proposed in a toy model by Haldane in 1988,[3] it took two and a half decades until it was implemented in real materials, Cr-doped (Bi, Sb)$_2$Te$_3$ (CBST) films.[4] Nonetheless, the fully quantized QAHE was observed only at extremely low temperatures (< 100 mK) under an external electric field. This is a serious obstacle not only to practical applications but also to fundamental studies, because a multitude of probes are difficult to utilize below ~2 K and/or are not compatible with the field effect gating geometry. It is generally believed that disorder inherent in the doping process is responsible for the extremely low temperature required for QAHE ($T_{QAHE}$),[4-6] and several approaches have been made to boost $T_{QAHE}$. One is to mix different (Cr and V) magnetic dopants.[7] Second is to use a magnetic modulation doping scheme, where high density of magnetic dopants are placed close to the surfaces,[5] and another is to utilize magnetic proximity effect to suppress the disorder.[8-10] There have also been efforts to implement high temperature QAHE in other materials platforms, such as antiferromagnetic topological insulator MnBi$_2$Te$_4$,[11] twisted bilayer graphene,[12] and transition metal dichalcogenide MoTe$_2$/WSe$_2$ bilayer.[13] However, these materials are limited to microscopic flakes and suffer from incomplete quantization and reproducibility problems. Moreover, in these platforms, QAHE still remains in the sub-K regime without electric-field-effect gating, and even with gating, QAH signatures disappear well below 10 K.

Another obstacle of QAHE to its broader applications is its instability against exposure to air. It is well known that the Fermi level ($E_F$) of topological insulators (TIs) shifts substantially when they are exposed to the atmosphere [14] due to surface reaction with oxygen and water molecules. This effect is amplified for the QAHE in magnetic topological insulators (MTIs) due to the requirement that $E_F$ should be within the very small magnetic gap at the Dirac point.[15, 16] In order to utilize QAHE for wider applications, it is also critical to find a way to suppress the air-degradation problem, and the only viable way to solve this is to utilize some capping layer. So far researchers have successfully utilized two capping layers, Te and amorphous AlO$_x$ ($a$-AlO$_x$), for the BST system. Between these, $a$-AlO$_x$ is found to be more effective, presumably due to its much larger energy gap than that of Te.[16-18] Nonetheless, both Te and $a$-AlO$_x$ are passive capping layers, in the sense that their role is just limited to protection against reaction with air. However, we note the possibility that an amorphous CrO$_x$ ($a$-CrO$_x$) can be utilized as an active capping layer that can not only provide protection against air reaction, but also boost QAHE of the underlying CBST layer. Thin $a$-CrO$_x$ layer has been one of the most widely used protective layers around us, which makes the stainless-steel stainless. In addition, considering that Cr is the very magnetic dopant being used for the CBST platform, Cr would be a much more natural choice for the capping layer than Al.

We present in Figure 1 the main results of how the optimally chosen $a$-CrO$_x$ capping layer provides significantly enhanced QAHE and will discuss the details of how we arrived at these structures further below. In the optimal platform shown in Figure 1a, a very thin $a$-CrO$_x$ capping was prepared by depositing 0.45 nm-thick Cr at room temperature, in a molecular beam epitaxy (MBE) chamber, which then naturally oxidized in air after being taken out of vacuum. As for the MTI platform, we prepared a magnetic modulation-doped tri-layer heterostructure composed of lightly Cr-doped (Bi, Sb)$_2$Te$_3$ (C$_L$BST, 4 QL) layers sandwiched by heavily Cr-doped (Bi, Sb)$_2$Te$_3$ (C$_H$BST, 3 QL) layers (343) structure, on Al$_2$O$_3$(0001) substrate as depicted in Figure 1a. The magnetic modulation doping scheme is known to boost the QAHE performance by reducing the magnetic disorder in the main TI layer by separating active TI layers from heavily magnetic doped TI outer layers.[5] In our modulation doping scheme, we put a small, yet non-zero level of Cr dopant in the middle layer, and this turns out to be necessary to lock the coercive fields of the top and bottom ferromagnetic layers, as detailed below in Figure 3. Finally, we also utilized a few nanometer-thick epitaxial Cr$_2$O$_3$ as a buffer layer to help the CBST layer stick onto the Al$_2$O$_3$ substrate. As shown in Figure 4, the fact that we can achieve similar QAHE on a mica substrate without the buffer layer, suggests that, unlike the amorphous CrO$_x$ capping layer, the epitaxial (and antiferromagnetic) Cr$_2$O$_3$ buffer layer does not seem to play a significant role in boosting QAHE.

Figure 1a shows a magnetic modulation-doped tri-layer heterostructure composed of 4 QL of C$_L$BST layers sandwiched between 3 QL of C$_H$BST layers on Al$_2$O$_3$(0001) substrate, together with a sharp streaky RHEED pattern recorded before the $a$-CrO$_x$ capping layer. This displays the two-dimensional, epitaxial nature of the MTI stack. The Hall conductance ($\sigma_{xy}$) and longitudinal conductance ($\sigma_{xx}$) as a function of magnetic field ($B$) of sample #1, Cr$_{0.48}$(Bi$_{0.13}$Sb$_{0.87}$)$_{1.52}$Te$_3$/Cr$_{0.1}$(Bi$_{0.143}$Sb$_{0.857}$)$_{1.9}$Te$_3$/Cr$_{0.48}$(Bi$_{0.13}$Sb$_{0.87}$)$_{1.52}$Te$_3$ with 0.45 nm-thick Cr capping layer, is shown in Figure 1b and 1c. After cooling down the sample to 2.0 K - a temperature easily accessible on pumped liquid helium - with zero magnetic fields, $\sigma_{xy}(B)$ is measured while sweeping the $B$ field between 9 T and – 9 T. The $\sigma_{xy}$ reaches the fully quantized value of 1.00 $e^2/h$, where $e$ is the electron charge and $h$ is the Plank's constant, passing through the magnetic coercivity field ($B_c$) of 0.23 T. The fully quantized $\sigma_{xy}$ maintains its value through the maximum magnetic field of 9 T and its polarity changes from $e^2/h$ to $-e^2/h$ near $-B_c$. Overall $\sigma_{xx}$ remains minute (< 0.2 $e^2/h$), except the sharp peaks near $\pm B_c$, which originate from the formation of a multidomain state.[5, 19] The perfect ferromagnetic ordering of this film at 2.0 K can be verified by $B$-independent quantized $\sigma_{xy}$ of $e^2/h$.[4] The inset of Figure 1b shows the van der Pauw measurement geometry on the 1 cm × 1 cm sample with the four indium contacts at the corners.

One of the critical quantities determining the quality of a QAH sample is the size of the magnetic energy gap near the Dirac point. If the Fermi level of the QAH sample sits within this magnetic gap, the longitudinal conductance ($\sigma_{xx}$) should follow the Arrhenius equation of $\sigma_{xx} \propto \exp(-T_0/T)$, where $T_0$ gives a quantitative measure of the magnetic energy gap.[20] Nonetheless, first generation of the QAH samples did not exhibit such an activation behavior, and instead showed much slower, variable-range hopping conduction, indicating high level of disorders dominating the electric conductance.[4, 21] Later, improved quality films started showing activation behaviors.[7, 20, 22-24] In Figure 1d, we present the activation gaps of our 44 samples of various qualities against their Hall conductance values measured at 2.0 K ($\sigma_{xy, 2K}$). It shows a superlinear relationship between $T_0$ and $\sigma_{xy, 2K}$ with the maximum $T_0$ of 3.1 K, confirming strong correlation between $T_0$ and the quality of QAHE.

Figure 2a shows the temperature dependence of $\sigma_{xy}$(red) and $\sigma_{xx}$(black) monitored under $B$ = 0.01 T in sample #2, which is from a different batch but grown at the same growth condition as sample #1. With decreasing temperature, $\sigma_{xy}$ increases due to ferromagnetic ordering and a hysteresis loop emerges at 90 K. As the temperature decreases further, the Hall conductance becomes larger than longitudinal conductance, $\sigma_{xy}/\sigma_{xx} > 1$, below 15 K. $\sigma_{xy}(B)$ at various temperatures are shown in Figure 2b. $\sigma_{xy}(0)$ remains as high as 0.98 $e^2/h$ at 4.2 K, the liquid helium temperature at ambient pressure (red) and 0.92 $e^2/h$ at 10 K (green). It is also notable in Figure 2a (and in Figure S1 for comparison, supporting information) that the ferromagnetic transition temperature ($T_{FM}$), defined as the point when $\sigma_{xy}$ starts to rise above zero, is about twice higher in the $a$-CrO$_x$-capped samples than in the bare sample: 90 K vs 45 K, regardless of the $\sigma_{xy}$ (2.0 K) values. These results indicate that the $a$-CrO$_x$-capping layer does strengthen the ferromagnetic order of the MTI stack, which subsequently helps enhance the QAHE.

We will now discuss the details of how we arrived at this. Considering that magnetic proximity is known to provide magnetic coupling in AHE or QAHE systems[8, 9, 25] and that capping layer is required for long-term stability of the QAHE samples,[17] it would be most ideal if we can find an insulating capping layer that not only protects QAHE but also provides some magnetic proximity effect to boost its ferromagnetism without introducing extra disorder. This suggests that the best capping layer should be a ferromagnetic insulator (FMI) with perpendicular magnetic anisotropy: FMI with in-plane anisotropy or antiferromagnetic insulator can even harm QAHE by suppressing the perpendicular magnetic anisotropy of the QAHE platform. Unfortunately, most ferromagnetic materials are metallic and it is hard to find a FMI that can be deposited on top of the QAHE platform without degrading QAHE. However, we note that Cr, which is the very magnetic dopant of the CBST system, is known to exhibit various forms of magnetism. Pure Cr is an antiferromagnetic metal, CrTe is a ferromagnetic semimetal, $CrO_2$ is a ferromagnet metal, $Cr_2O_3$ is an antiferromagnetic insulator,[26-29] and $a$-$CrO_x$, which naturally forms when Cr is exposed to air, is a stable paramagnetic insulator that makes stainless-steel stainless by preventing it from being further oxidized. Moreover, there are even reports of ferromagnetism in ultrathin layers or nanoparticles of $a$-$Cr_2O_3$.[30] This suggests that thin amorphous $a$-$CrO_x$, being either paramagnetic with high magnetic susceptibility or even ferromagnetic, could enhance the ferromagnetic order of the underlying CBST stack without introducing any additional disorder.

In order to see the possibility of using $a$-$CrO_x$ as a capping layer for the CBST stack, we first compared the transport properties of MTIs with various capping layers in Figure 3a. For this study, we have chosen $(Cr_{0.1}Sb_{0.9})_2Te_3$ as the MTI layer, instead of the more complex 343 structure shown in Figure 1a, and compared four different capping layers: Te (40 nm), Cr (0.5 nm), $In_2O_3$ (15 nm) and CoO (5 nm), all deposited at room temperature. Oxide capping layers were grown by depositing each metal element in varied molecular oxygen pressures of $6 \times 10^{-7} - 2 \times 10^{-6}$ Torr (see details in the Experimental section). The thickness of each capping layer was calibrated with an in-situ quartz crystal microbalance (QCM) measurement. With decreasing temperature, $In_2O_3$-capped film becomes highly resistive. On the other hand, the CoO-capped film becomes highly conducting. This suggests that neither of these two materials can be used as a capping layer for the CBST system. On the other hand, the naturally oxidized Cr(i.e. $a$-$CrO_x$)-capped film is comparable to the Te-capped film, suggesting the possibility that thin $a$-$CrO_x$ layer may work as a good capping layer.

Before moving on to optimizing the growth of the *a*-CrO$_x$ as the capping layer, we first optimized the composition of the main modulated CBST stack. For this purpose, we first searched for the most insulating Cr-composition of the CBST system, to use it as the outer layers of the modulated CBST stack. As shown in Figure S2, CBST is most insulating near the Cr concentration of 0.45. Accordingly, we have chosen Cr ≈ 0.45 as the outer layers of the modulated CBST stack. Then, testing on various modulation structures, including 232, 242, 252, 333, 343, and 353 with an identical growth recipe, we find that 343 shows the best performance (Figure S3, supporing information). Then, we compared two 343 CBST stacks, one with a very thin *a*-CrO$_x$ as the capping layer and the other with Te as the capping layer in Fig. 3b. This shows drastic improvement of both $\sigma_{xy}$ and $\sigma_{xx}$ with the *a*-CrO$_x$ capping layer: $\sigma_{xy}$ increases more than twice while $\sigma_{xx}$ decreases ~7 times at 2.0 K compared to those of the passive Te-capped sample. It is also notable that *a*-CrO$_x$-capped system exhibits two coercive fields, while Te-capped system exhibits only one, which sits between the two coercive values of the *a*-CrO$_x$-capped film. The fact that the dual coercive fields are observed only in the *a*-CrO$_x$-capped film suggests that the *a*-CrO$_x$-capping layer enhances the ferromagnetism of only the top layer, but not the bottom layer.

Since we confirmed that *a*-CrO$_x$ capping layer can boost QAHE through enhanced ferromagnetism, we studied its thickness effect in Figure 3c. We compared three different thicknesses of Cr-capping, 0.3, 0.4, and 0.5 nm: best QAHE samples should exhibit minimal $\sigma_{xx}$ and maximal $\sigma_{xy}$ (as close as possible to $e^2/h$). 0.4 and 0.5 nm-thick Cr-capped films show comparably low $\sigma_{xx}$ at 2 K, while 0.3 nm sample exhibits $\sigma_{xx}$ larger than $e^2/h$. $\sigma_{xy}$ (*B*) curves also show consistent behaviors: $\sigma_{xy}$ of 0.4 and 0.5 nm samples reaches 0.8 $e^2/h$ at 1 T, while 0.3 nm Cr sample exhibits much lower ~0.6 $e^2/h$ of $\sigma_{xy}$. Overall, 0.4 nm film shows the best behaviors in both $\sigma_{xy}$ and $\sigma_{xx}$. It is notable that 0.3 nm film does not show the dual coercive fields. All these results suggest that thinner than 0.4 nm Cr capping layer is too thin to fully cover the entire MTI surface and does not provide the ferromagnetism-boosting proximity effect, whereas thicker than 0.5 nm capping layer has some metallic Cr layer left, not fully oxidized after exposure to air. Despite these improvements created by the Cr (*a*-CrO$_x$) capping layer, this C$_H$BST/BST/C$_H$BST structure has a serious problem of dual coercive fields with the smaller value too close to the zero magnetic field [18]. So, we need further improvement of the structure to solve this.

We were able to solve this dual coercive field problem by introducing a small concentration of Cr-doping in the middle layer, which provides magnetic coupling between the top and bottom layers: Figure S4 shows that the double-step behavior gradually diminishes with increasing Cr-doping in the middle layer and completely disappears with Cr = 0.1 doping. With all these optimized building blocks, we finally observed the fully quantized $\sigma_{xy}$ at 2 K in Figure 3d: black, red, and green curves indicate samples capped with different thickness of Cr capping layers: 0.40, 0.45, and 0.50 nm, respectively. Within this range of Cr thickness variation, there is no noticeable difference in $\sigma_{xy}(B)$.

Next, in order to see how these samples degrade in air, we measured, in Figure 4a, samples #4 and #5 grown with an identical recipe except for capping layers, over an extended period of time. We deposited 0.4 nm-thick Cr capping only on sample #5. Sample #5 shows quantized value of $\sim e^2/h$ at 2 K, while $\sigma_{xy}$ of a bare sample, sample #4, is only $1/3\ e^2/h$. Figure 4a compares how Hall conductance degrades in bare (black square) and $a$-CrO$_x$-capped (green triangle) samples. The degradation - quantified as $\delta\sigma_{xy} = [(\sigma_{xy}(d) - \sigma_{xy}(0)]/\sigma_{xy}(0)$, where $d$ is the number of days from fabrication - of the bare sample is nearly 45 % over 13 days. On the other hand, Cr-capped (green triangle) sample shows less than 20 % of degradation over 38 days. With the thin $a$-CrO$_x$ as the capping layer, Hall conductance increases by 3 times and stability is also enhanced by 4 times. The protection offered by $a$-CrO$_x$ capping against air is comparable or slightly better than that of AlO$_x$ passive capping layer.[17] In order to improve the stability even further, we prepared another sample (sample #6, red circles in Fig. 4A) and applied additional, 7 nm-thick, Se-capping layer on top of the $a$-CrO$_x$-capping layer. Then, the degradation is further reduced by 12 times, from 19 % to 1.6 % over the same duration of 38 days: the inset shows the hysteresis loop of $\sigma_{xy}(B)$ for this sample, as grown (black) and after 76 days (red).

Finally, we also employed a flexible mica substrate, and achieved similar QAHE as shown in Figure 4b. First of all, this demonstration implies that the epitaxial Cr$_2$O$_3$ buffer layer, which was used to improve sticking on the Al$_2$O$_3$ substrates, is not essential to boosting QAHE, because the Cr$_2$O$_3$ buffer layer was not used for the mica substrate. High-temperature QAHE on flexible mica substrates also opens the door to numerous applications. First, mica is much more economical than any other substrates, such as GaAs, InP, and SrTiO$_3$, previously used for QAHE.[4, 5, 7, 20, 31] Second, being a van der Waals material, mica could be easily thinned down for effective backgating.[4, 7] Third, mica substrate, being flexible, would allow easy implementation of strain effect study on QAHE.[32, 33] In order to demonstrate this, we transferred the QAHE film and the mica substrate to a thin transparent plastic as shown in the photo of Figure 4b and carried out bending tests. The transferred QAHE sample did not exhibit any signature of cracks under an optical microscope after several bending tests, suggesting robustness of this platform for strain studies.

In summary, utilizing a thin amorphous $CrO_x$ as an active magnetic capping layer on an optimally modulated CBST stack, we achieved substantially enhanced QAHE for both $T_{QAHE}$ and long-term stability, on centimeter-scale, sapphire as well as flexible mica substrates, which opens the possibility of strain studies on QAHE. Foremost, the $a$-$CrO_x$ capping layer alone doubled the ferromagnetic Curie temperature of the underlying CBST layer from 45 to 90 K, likely through magnetic proximity effect, and more than doubled the Hall conductance. Then, with additional heterostructure engineering schemes, we achieved quantized Hall conductance, 1.00 $e^2/h$ at 2.0 K, 0.98 $e^2/h$ at 4.2 K and 0.92 $e^2/h$ at 10 K without electric gating. With an additional passive capping layer, the high-temperature QAHE was maintained over a months' time scale. All in all, this work would allow QAHE to be accessible to a broad family of probes that have been out of reach for QAHE studies due to technical barriers such as extremely low $T_{QAHE}$, electric gating requirement, microscopic sample size, expensive non-versatile substrates, and aging effects, and may pave a way to practical quantum anomalous Hall devices.

**Materials and Methods**

**Thin-film growth.** Films were grown on 1 cm × 1 cm $Al_2O_3$ (0001) and flexible mica (Ted Pella) substrates using a custom-built MBE system (SVTA) with base pressure less than $5 \times 10^{-10}$ Torr. Substrates were cleaned ex-situ by UV-generated ozone and in situ by heating to 750 °C under oxygen pressure of $1 \times 10^{-6}$ Torr. High-purity (99.999%) Bi, Sb, Te, Cr, and Co were thermally evaporated using Knudsen effusion cells for the film growth. All the source fluxes were calibrated in-situ by quartz crystal micro-balance and ex-situ by Rutherford backscattering spectroscopy. All capping layers were deposited at room temperature on top of the films: Te (40 nm, flux of $1 \times 10^{14}$ $cm^{-2} \cdot sec^{-1}$), Cr (0.5 nm, flux of $1 \times 10^{13}$ $cm^{-2} \cdot sec^{-1}$), $In_2O_3$ (15 nm, flux of $1 \times 10^{13}$ $cm^{-2} \cdot sec^{-1}$ in $2 \times 10^{-6}$ Torr of molecular oxygen), and CoO (5 nm, flux of $2.9 \times 10^{13}$ $cm^{-2} \cdot sec^{-1}$ in $6 \times 10^{-7}$ Torr of molecular oxygen).

**Transport measurements.** All transport measurements were performed using the standard van der Pauw geometry in centimeter-scale samples, by manually pressing four indium wires on the corners of each sample. Keithley 2400 source-measure units and 7001 switch matrix system were controlled by a LabView program for the van der Pauw measurement. Advanced Research Systems closed-cycle cryostat and an electromagnet were used for magneto-transport measurements down to ~6 K. Quantum anomalous Hall effect measurements were carried out in a Physical Property Measurement System (PPMS, Quantum Design inc.) down to 2 K.

**Supplementary information**

Figs. S1 to S4 for control tests

**Acknowledgements**

This work is supported by National Science Foundation's DMR2004125, Army Research Office's W911NF2010108, MURI W911NF2020166, and the center for Quantum Materials Synthesis (cQMS), funded by the Gordon and Betty Moore Foundation's EPiQS initiative through Grant GBMF10104. We would like to thank Molly P. Andersen, Linsey K. Rodenbach, and David Goldhaber-Gordon at Stanford University for discussions and suggesting evaluation of the activation energies, and Yoshinori Tokura at University of Tokyo for helpful discussions.

**Author Contributions**

S.O. and H.T.Y. conceived the experiments. H.T.Y., D.J., and X.Y. performed thin film growth and H.T.Y. performed transport measurements. S.O. and H.T.Y. analyzed the data and wrote the paper. All authors discussed the results.

**Conflict of Interest:**

The authors declare no conflict of interest.

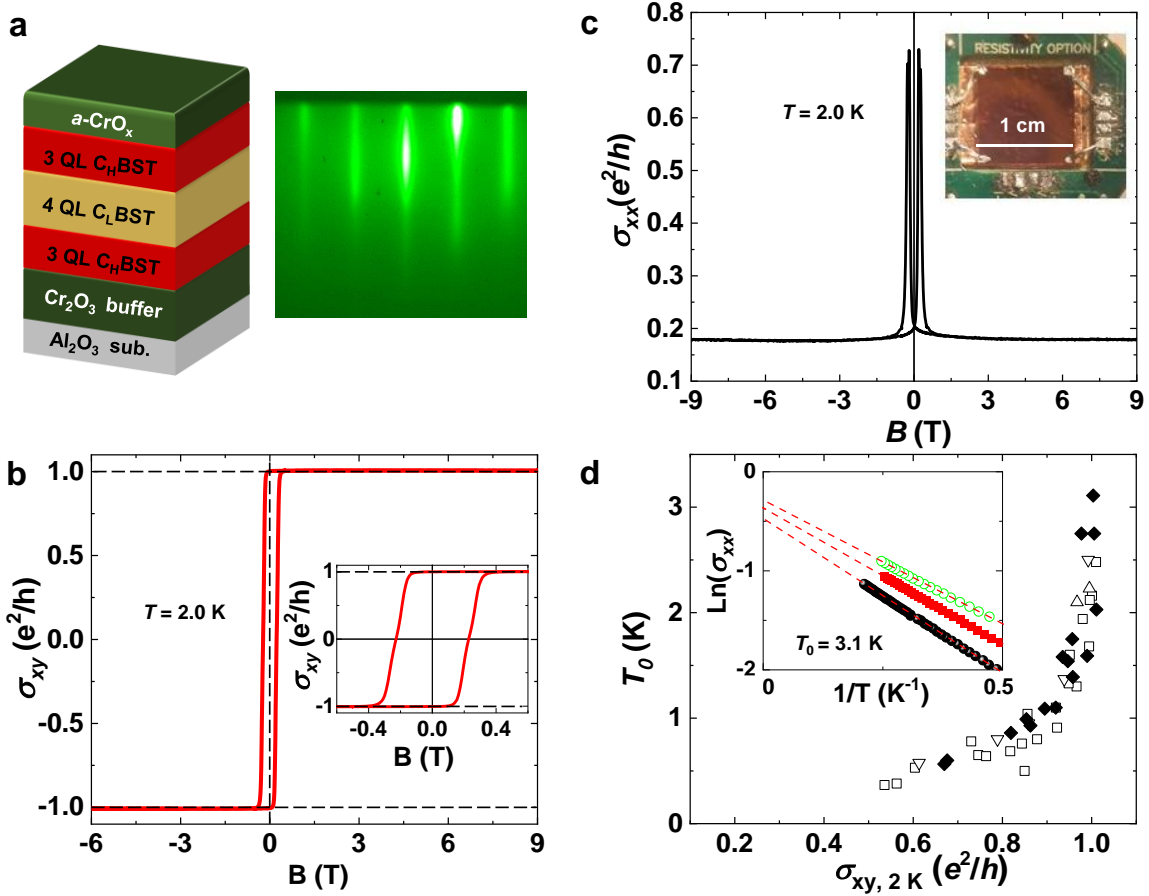

**Figure 1.** High-temperature QAHE platform and the transport properties. a) Schematic layout of the tri-layer Cr-doped $(Bi_xSb_{1-x})Te_3$ MTI structure, sandwiching lightly Cr-doped BST(4 QL) layer between two heavily Cr-doped BST(3 QL) layers and RHEED pattern of a CBST film b) Fully quantized Hall conductance ($\sigma_{xy}$) for sample #1, monitored at 2.0 K while sweeping the magnetic field between ±9 T. c) $\sigma_{xx}(B)$, exhibiting low conductance of < 0.2 $e^2/h$ except near coercivity fields. The inset shows the 1 cm × 1 cm size sample with four indium contacts configured for the van der Pauw geometry of transport measurement. d) Activation energy gaps as a function of $\sigma_{xy,\,2\,K}$ for 44 samples of various qualities, measured at multiple magnetic fields of 0.01 (squares), 0.1 (up-pointing triangles), 0.2 (down-pointing triangles), and 9 T (diamonds) while warming. The inset shows the Arrhenius fitting of sample #1 ($T_0 = 2.7$ K, open green circles), sample #2 ($T_0 = 2.5$ K, red solid squares), and sample #3 ($T_0 = 3.1$ K, black solid circles). Red dashed lines indicate the fitting lines with the Arrhenius equation $\sigma_{xx} \propto \exp(-T_0/T)$.

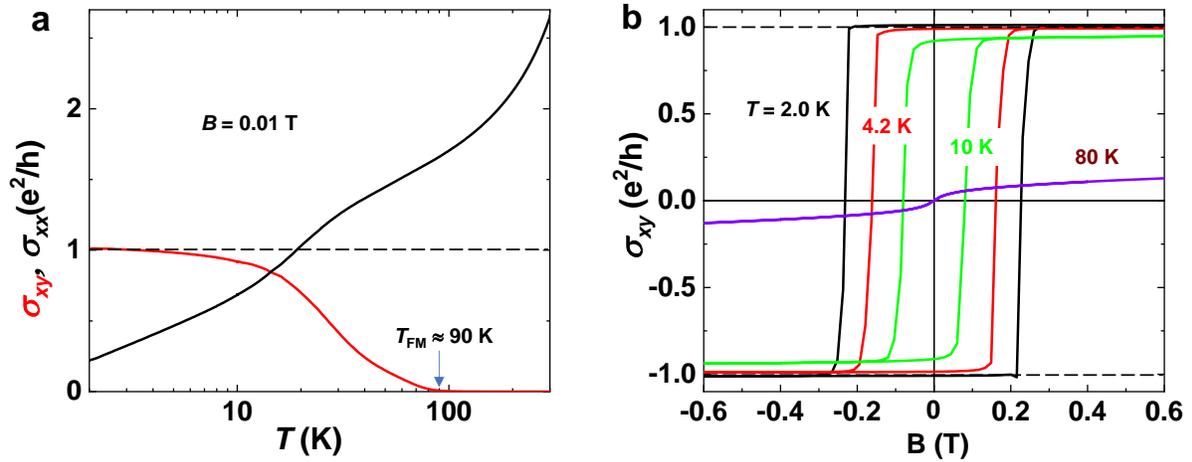

**Figure 2.** Temperature dependence of Hall and longitudinal conductances. a) $\sigma_{xx}$ (black) and $\sigma_{xy}$ (red), monitored during the warming after positive poling under magnetic field of $B = 0.01$ T for sample #2. b) Magnetic field dependence of Hall conductance for sample #2 at various temperatures of 2.0 (black), 4.2 (red), 10 (green), and 80 K (violet). Note that $\sigma_{xy}(0)$ maintains 0.98 $e^2/h$ at 4.2 K and 0.92 $e^2/h$ even at 10 K.

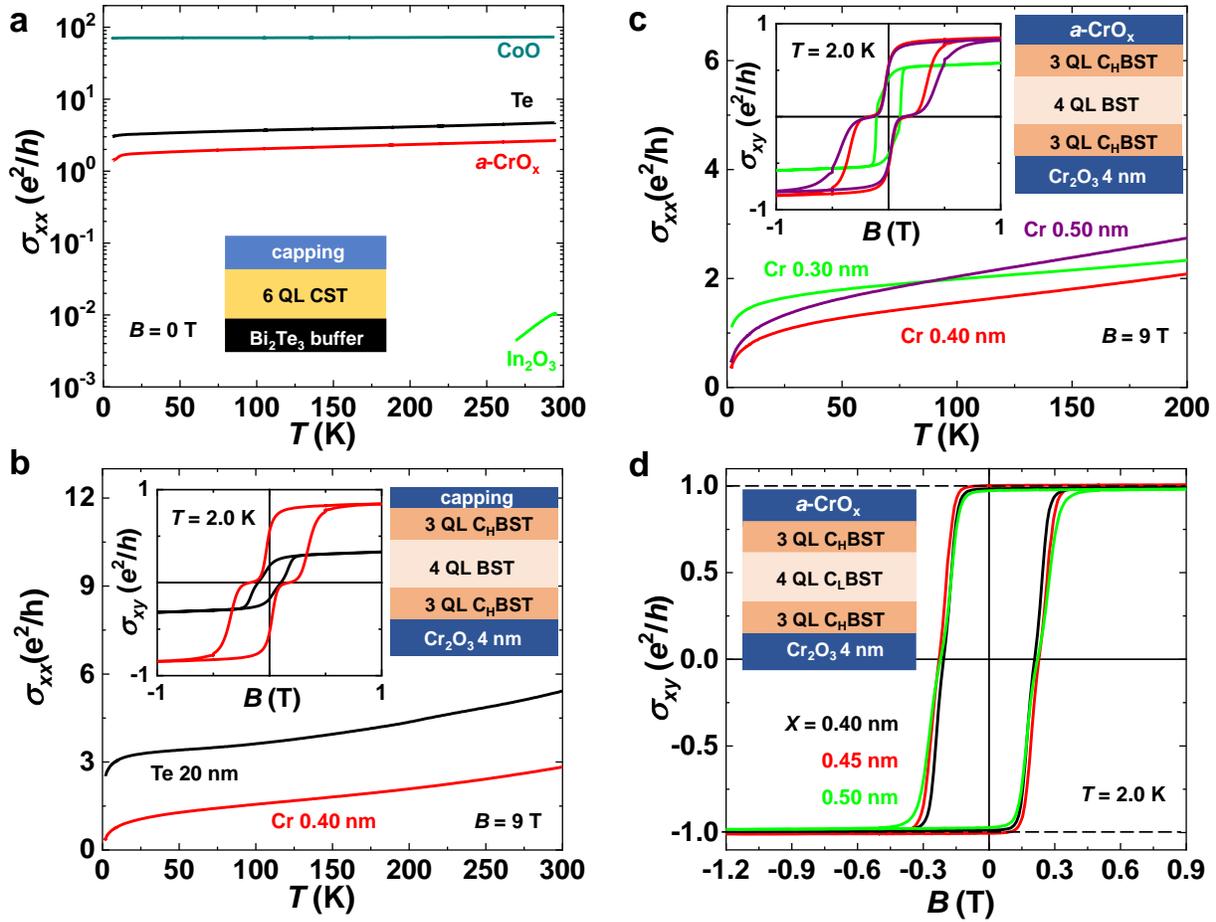

**Figure 3.** Schematics and transport properties of various control samples. a) Temperature dependence of $\sigma_{xx}$ for various capping materials: CoO (5 nm, dark green), Te (40 nm, black), Cr (0.50 nm, red), and In$_2$O$_3$ (15 nm, green). Cr is fully oxidized into a-CrO$_x$ when exposed to air. Comparison of $\sigma_{xx}(T)$ and $\sigma_{xy}(B)$ for b) Te (black) and Cr (red, 0.40 nm) capped samples, and c) different thicknesses of Cr capping layer of 0.30 (green), 0.40 (red), and 0.50 nm (purple), respectively. d) Black, red, and green curves indicate $\sigma_{xy}(B)$ for various Cr thicknesses of 0.40, 0.45, and 0.50 nm, respectively: they are all comparable, but 0.45 nm sample shows slightly better $\sigma_{xy}(B)$ among them.

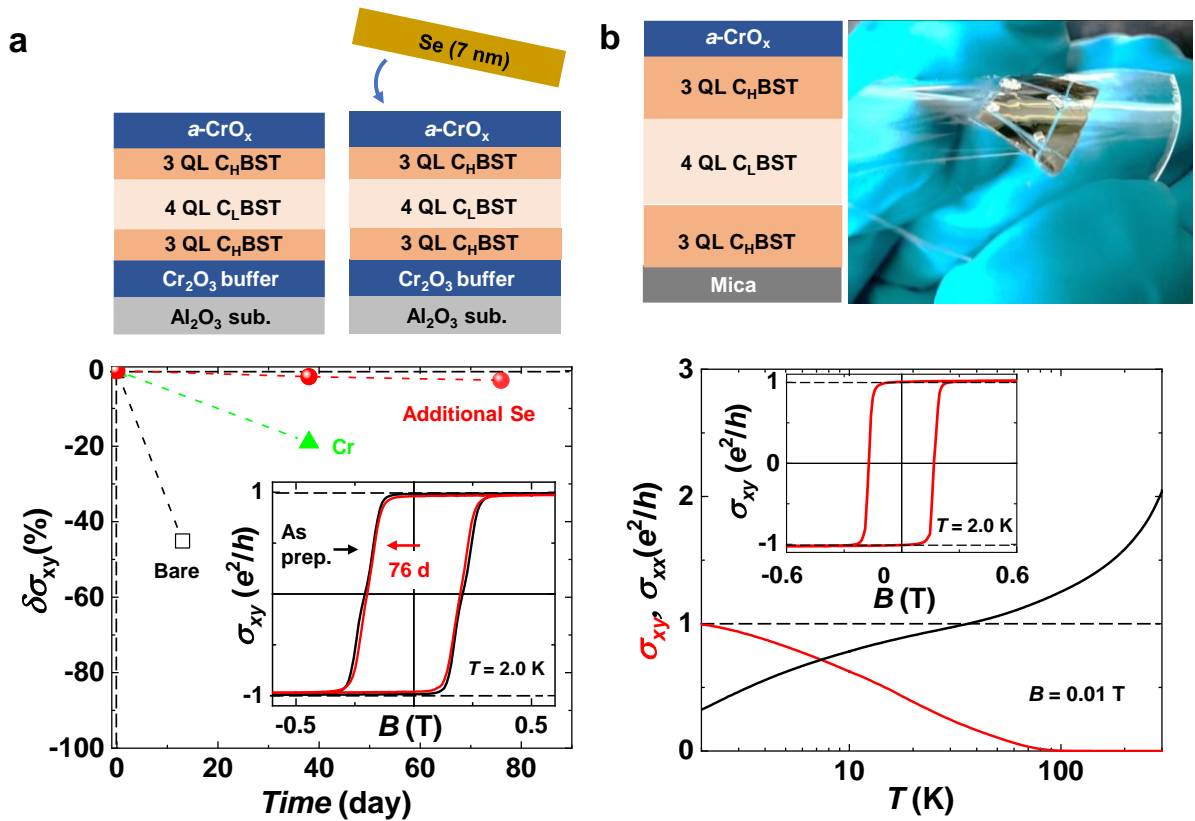

**Figure 4.** Aging tests and QAHE film on a flexible mica substrate. a) Hall conductance at $B = 0$ T of sample #4 (black square) without capping layer, sample #5 with Cr capping (green triangle), and sample #6 with additional Se capping (red circle). Inset shows $\sigma_{xy}(B)$ of as-prepared (black) and measured 76 days after the first measurement (red) in sample #6. b) Schematic layout of the QAH film grown on a flexible mica substrate without the $Cr_2O_3$ buffer layer. The QAH film on the mica substrate was attached to a thin plastic to demonstrate bendability. The bottom shows the temperature dependence of $\sigma_{xx}$ (black) and $\sigma_{xy}$ (red) of this film measured at $B = 0.01$ T, with the inset showing the $\sigma_{xy}(B)$ curve.

Supporting Materials

**High temperature, gate-free quantum anomalous Hall effect with an active capping layer**

*Hee Taek Yi[1,2], Deepti Jain[1], Xiong Yao[2†], and Seongshik Oh[1,2]\**

[1] Department of Physics and Astronomy, Rutgers, The State University of New Jersey, Piscataway, NJ 08854, USA.
[2] Center for Quantum Materials Synthesis, Piscataway, NJ 08854, USA.
†Present address, Ningbo Institute of Materials Technology and Engineering, Chinese Academy of Sciences, Ningbo 315201, China.

\*Corresponding author. Email: ohsean@physics.rutgers.edu

*Control test 1: Comparison of ferromagnetic onset temperatures with and without a-$CrO_x$ cap*

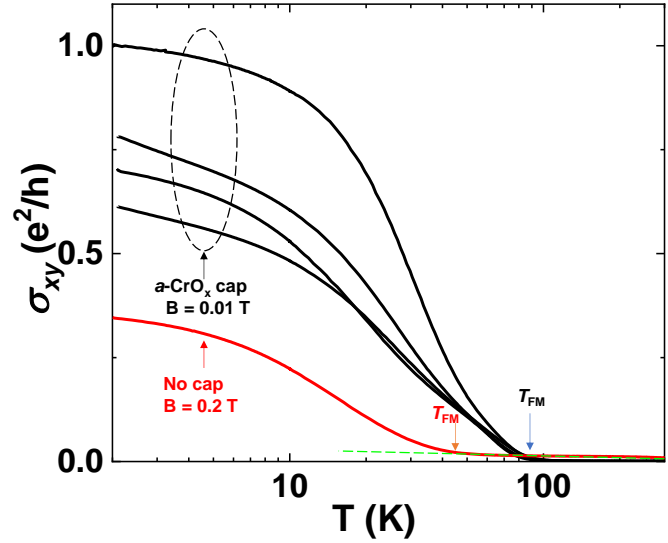

**Figure *S1*.** Temperature dependence of Hall conductance with and without *a*-$CrO_x$ cap. Red and black curves indicate the $\sigma_{xy}(T)$ for bare and *a*-$CrO_x$ capped samples, respectively. Ferromagnetic ordering occurs at 45 K for the bare sample and at 90 K for the *a*-$CrO_x$-capped samples: note that all the *a*-$CrO_x$-capped samples exhibit similar $T_{FM}$ despite their quite different $\sigma_{xy}$(2 K) values. This implies two things. First, the main effect of the *a*-$CrO_x$ capping layer is boosting of the ferromagnetism. Second, even with the boosted ferromagnetism, other factors need to be optimized for QAHE. While the bare sample is measured at 0.2 T, all the *a*-$CrO_x$-capped samples are measured at 0.01 T. Accordingly, only the bare sample exhibits non-negligible ordinary Hall conductance, as indicated by the green dashed line.

*Control test 2: Search for an optimal Cr concentration for the outer ferromagnetic layer of the modulated tri-layer structure*

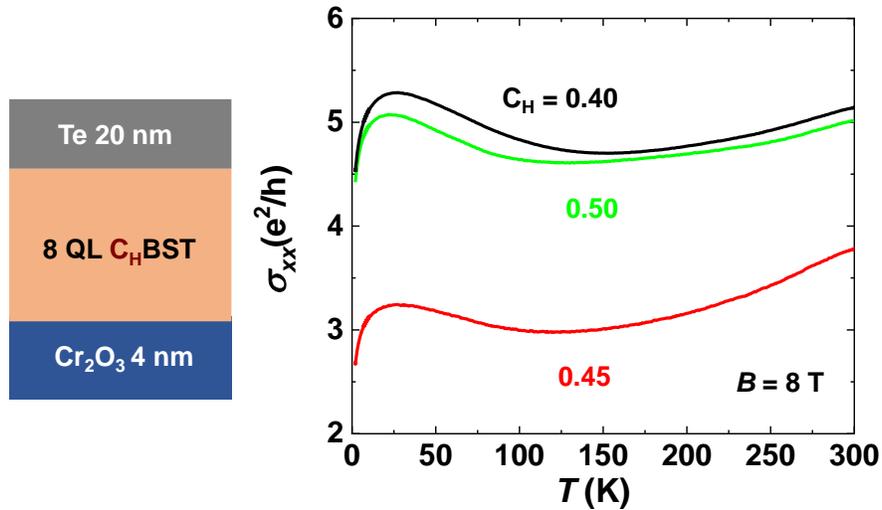

**Figure *S2*.** Longitudinal conductance as a function of temperature at various Cr concentrations searching for the least conductive CBST layer. For this purpose, we chose 8 QL $C_H$BST as the MTI layer with a tellurium capping layer. Black, red, and green curves indicate the Cr concentrations of 0.40, 0.45, and 0.50, respectively. CBST shows the least conducting (or most insulating) behavior near the Cr doping of 0.45.

*Control test 3: Search for the optimal thicknesses of outer and inner layers for the modulated tri-layer structure*

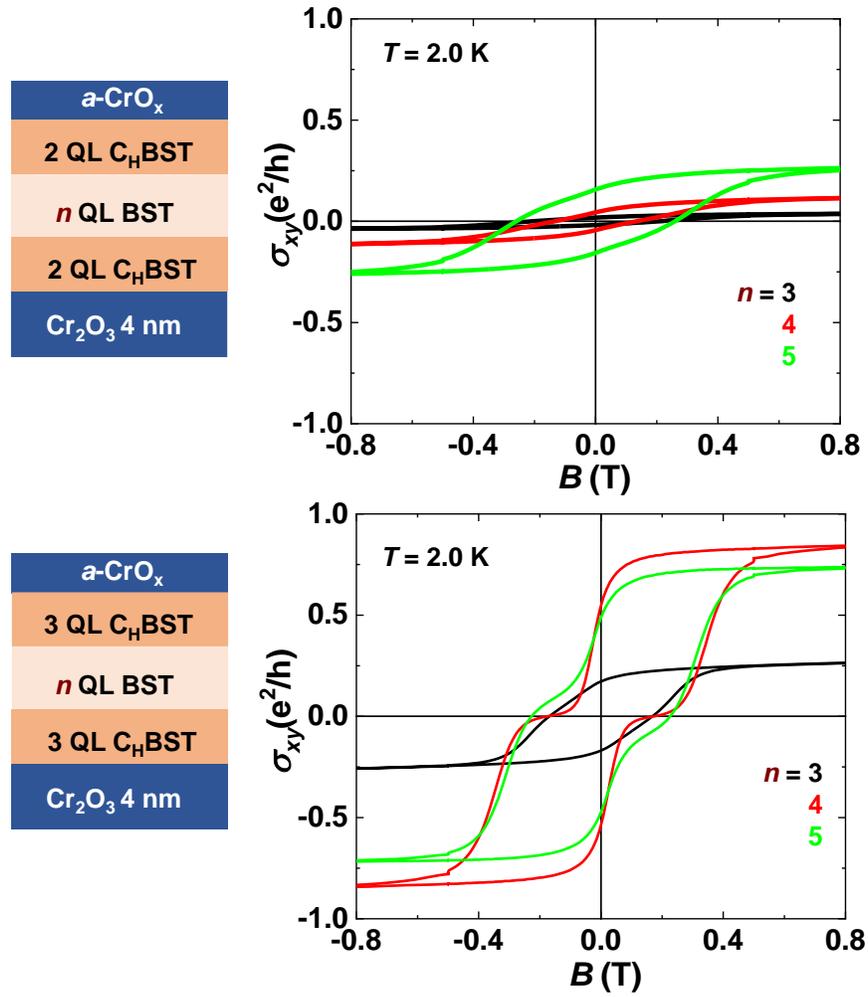

**Figure S3.** Layer thickness dependence of the Hall conductance at 2.0 K. The magnetic field dependence of Hall conductance at various inner and outer layer thicknesses shows that 343 structure provides the largest Hall conductance.

*Control test 4: Cr-doping of the inner layer to suppress the dual coercive fields, which finally achieved the high temperature QAHE*

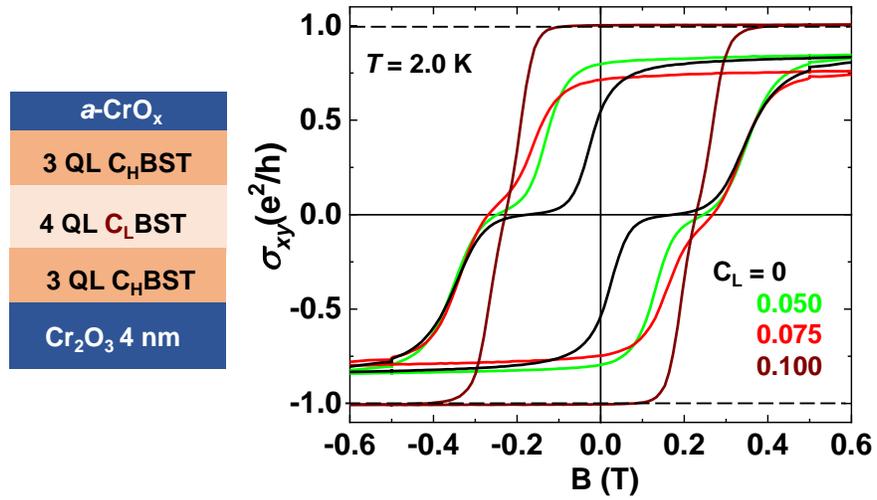

**Figure S4.** Hall conductance dependence on the Cr concentration ($C_L$) of the inner layer at 2.0 K. Black-, red-, green-, and brown curves indicate the $C_L$ of 0, 0.050, 0.075, and 0.100, respectively. The double-step feature gradually diminishes with increasing $C_L$ and totally disappears at $C_L$ = 0.1.